\documentclass[10pt]{article}
\usepackage{cite} 
\usepackage{graphicx,amsmath,amssymb,amsbsy,latexsym,amsthm}
\usepackage[mathscr]{eucal}

\theoremstyle{plain}

\def\C{\mathbb{C}}

\def\tr{\mathrm{tr}}

\newcommand{\SU}{\mathrm{SU}}

\newcommand{\eqa}{\begin{eqnarray}}
\newcommand{\neqa}{\end{eqnarray}}
\newcommand{\be}{\begin{equation}}
\newcommand{\ee}{\end{equation}}

\newcommand{\Hil}{\mathcal{H}}

\newcommand{\CP}{\mathbb{CP}}

\newcommand{\etan}{\hat{\eta}}
\newcommand{\etap}{\zeta}

\newcommand{\toconsider}[1]{}

\newcommand{\fixedj}{\mathfrak{j}} 

\newcommand{\critpt}{\text{crit. pt.}}
\newcommand{\crit}{\text{crit}}
\newcommand{\criteq}{\underset{x_o}{=}}

\newcommand{\SEPRLorig}{S} 
\newcommand{\Sfinal}{\tilde{S}^o} 
\begin{document}

\title{Hessian and graviton propagator of the proper vertex}
%
%
\author{
Atousa Chaharsough Shirazi\thanks{achahars@fau.edu},
Jonathan Engle\thanks{jonathan.engle@fau.edu}, and
Ilya Vilensky\thanks{ivilenskiy2013@fau.edu}\\
\\
{\it Department of Physics, Florida Atlantic University} \\
{\it 777 Glades Road, Boca Raton, FL 33431, USA }
}

\maketitle

%
%

\begin{abstract}

The proper spin-foam vertex amplitude is obtained from the EPRL vertex by projecting out all but a single gravitational sector, in order to achieve correct semi-classical behavior. In this paper we calculate the gravitational two-point function predicted by the proper spin-foam vertex to lowest order in the vertex expansion.  We find the same answer as in the EPRL case
 in the `continuum spectrum' limit,
 so that the theory is consistent with the predictions of linearized gravity in the regime of small curvature.
The method for calculating the two-point function is similar to that used in prior works:
we cast it in terms of an action integral and to use stationary phase methods.  Thus, the calculation
of the Hessian matrix plays a key role.  Once the Hessian is calculated, it is used not only to calculate the two-point function,
but also to calculate the coefficient appearing in the semi-classical limit of the proper vertex amplitude itself.  This coefficient
is the effective discrete ``measure factor'' encoded in the spin-foam model.
Through a non-trivial cancellation of different factors, we find that this coefficient is the same as the
coefficient in front of the term in the asymptotics of the EPRL vertex corresponding to the selected gravitational sector.

\end{abstract}

\section{Introduction}

Loop quantum gravity (LQG) is an approach to the canonical quantization of general relativity (GR) that takes as
its guiding principle that there is no background space-time geometry, but rather that gravity is geometry.  The path integral approach to the dynamics of LQG leads to the
\textit{spin-foam} framework.
The currently most used spin-foam model of the dynamics of LQG is the so-called
\textit{Engle-Pereira-Rovelli-Livine} (EPRL) model \cite{fk2007, elpr2007, kkl2009}.
One difficulty with this and all spin-foam models before it is that, due to the inclusion of more than a single gravitational sector
\cite{engle2011,clrrr2012},
solutions to the classical equations of motion of GR fail to dominate in the semi-classical limit \cite{engle2011a, engle2012}.
This problem was solved by quantum mechanically imposing restriction to a single gravitational sector,
yielding what has been called the `proper' spin-foam model of quantum gravity \cite{engle2011a, engle2012}.

However, this revised model of quantum gravity, like any proposed model of quantum gravity, must pass certain tests.
For example, in cases where the space-time curvature is small, one expects linearized quantum gravity to be correct --
current experiments such as BICEP and Planck have the potential to test this expectation \cite{kw2014, BICEP22014}.
As a consequence, in this regime, one expects full
quantum gravity to reproduce the predictions of the linearized theory.
In the present paper, as in the prior papers \cite{bmp2009b, bd2011}, we specifically consider the \textit{two-point correlation
function} of the gravitational field, also referred to in the literature as the `graviton propagator'
\cite{rovelli2005, bmrs2006, ar2007, ar2007a, abr2008, bmp2009b, bd2011}.
The EPRL model was shown to reproduce the same two-point correlation function as linearized gravity
in prior work by Bianchi, Magliaro, Perini, and Ding \cite{bmp2009b, bd2011}.  In the present paper we show that this success of the EPRL model extends to the proper vertex.  This involves casting the appropriate amplitude in an integral form with an action, and finding the Hessian of the action.
We restrict consideration to the Lorentzian signature version of the model \cite{ez2015, evz2015, engle2012}, as this is the physically relevant one. 

For the propagator it turns out that, in the same asymptotic limit considered in \cite{bmp2009b, bd2011}, one obtains the same answer as for EPRL,
so that the proper vertex continues to be consistent with linearized gravity
to lowest order in the vertex expansion.  However, when more than one vertex (simplex) is considered, due to
the considerations in \cite{engle2012, engle2011a}, we expect the EPRL and proper vertex amplitudes to yield different propagators, with only the proper vertex being consistent with linearized gravity.
This will be investigated in future work and is touched upon briefly in the discussion section of this paper.

A further benefit of the calculation of this Hessian is that it permits an easy determination of the coefficient in
front of the semi-classical limit of the proper vertex amplitude derived in \cite{evz2015}, important for understanding the
effective `measure factor' of the model \cite{ce2013, eht2009}.
We show that this coefficient is exactly the same as the coefficient of the term in the semi-classical limit of the EPRL
vertex amplitude corresponding to the isolated gravitational sector.
This comes about through a non-trivial cancellation of the new factors in the determinant of the Hessian and in the integration measure for the additional variables present in the integral form of the proper vertex amplitude.

The paper is organized as follows.  In section \ref{sect:prelim}, the EPRL vertex amplitude, proper vertex amplitude, and graviton
propagator calculation for EPRL are briefly summarized.
In section \ref{sect:propagator}, the modifications to the strategy for calculating the
graviton propagator required for the proper vertex are presented.
In section \ref{sect:hess},
all new components of the Hessian not present in the original EPRL work are calculated.
In section \ref{sect:gammazero} the resulting expression for the propagator in the `continuum spectrum' limit is given.
Finally, in section \ref{sect:coeff}, the determinant of the Hessian is calculated, thereby providing the coefficient in the asymptotics of the proper vertex amplitude. We then close with a discussion of these results and future work.

\section{Preliminaries}
\label{sect:prelim}

In this section we briefly recall the prior constructions and results required for the present paper, and fix conventions.

\subsection{EPRL vertex}
\label{sect:EPRL}

We start by recalling the $SL(2,\mathbb{C})$ EPRL vertex amplitude,
defined on a given oriented 4-simplex.
The tetrahedra are numbered from $0$ to $4$
in a manner consistent with the fixed orientation of the 4-simplex in the sense defined in \cite{engle2011}.
The relevant boundary Hilbert space is spanned by $SU(2)$ spin networks
$\Psi$ labelled by spins $j_{ab}$ and vectors $\psi_{ab}$, $\psi_{ba}$ in the corresponding irreducible representation of $SU(2)$,
and given explicitly by $\Psi(g_a)=\sum\limits_{a<b}\langle\psi_{ab}|g_a^{-1}g_b|\psi_{ba}\rangle$,
where $a,b$ run from $0$ to $4$.

The principal series of irreducible representations $\Hil_{k,p}$ of $SL(2,\mathbb{C})$ are labelled by real numbers $p$ and half-integers $k$. Each such $SL(2,\mathbb{C})$ irreducible representation $\mathcal{H}_{k,p}$
can be decomposed into a direct sum of $SU(2)$ irreducibles $\mathcal{H}_j$: $\Hil_{k,p} = \oplus_{n=0}^\infty \Hil_{k+n}$.

The amplitude for a single 4-simplex is then given by:
\begin{align}
\label{EPRLvertexorig}
A_v = \int_{SL(2,\mathbb{C})^5} \delta(g_0) \prod_a dg_a \prod_{a<b} \mathcal{P}_{ab}
\end{align}
where
\begin{align*}
\mathcal{P}_{ab} = \alpha(\mathcal{I} \psi_{ab}, g^{-1}_a g_b \mathcal{I} \psi_{ba})
\end{align*}
is defined for each triangle $(ab)$. Here $\alpha$ is the invariant bilinear form satisfying
$\alpha(\psi, \phi)=(-1)^{2k}\alpha(\phi, \psi)$ \cite{bdfhp2009}.
The map $\mathcal{I}$ embeds the spin $j$ irrep into the lowest $SU(2)$ irrep in decomposition of
$\mathcal{H}_{k,\gamma k}$.
Using the \mbox{(skew-)symmetry} of $\alpha(\cdot, \cdot)$, it follows that the resulting
amplitude (\ref{EPRLvertexorig}) is independent of the choice of the tetrahedra,
so long as this numbering is consistent with the fixed 4-simplex orientation.

The spin $j$ representation of $SU(2)$ can be realized as the action of $SU(2)$ on degree $2j$ polynomials
of a spinor $z^A \in \C^2$ \cite{bdfhp2009, ez2015} where $A=0,1$. In terms of this realization, for each spin $j$ and spinor $\xi$,
one can define the coherent state $C^j_{\xi}$:
\begin{align*}
C^j_{\xi}(z) = \sqrt{\frac{d_j}{\pi}} \langle\bar{\xi},z\rangle^{2j}
\end{align*}
where  $\langle w,z \rangle = \bar{w}_0 z_0 + \bar{w}_1 z_1$ and $d_j=2j+1$.
Furthermore, every spinor $\xi$ is a coherent state associated with a vector $n_{\xi}$
defined by $(n_{\xi}\cdot\sigma)\xi = \xi$. The map $J$
\begin{align*}
 J \begin{pmatrix} \xi_0 \\ \xi_1 \end{pmatrix} =  \begin{pmatrix} -\bar{\xi}_1 \\ \bar{\xi}_0 \end{pmatrix}
\end{align*}
applied to $\xi$ yields a spinor satisfying $n_{J\xi} = -n_{\xi}$.

$\Hil_{k,p}$ can similarly be realized as a space of homogeneous functions of a spinor $z^A$.
Using the explicit expression for $\mathcal{I}$ and $\alpha(\cdot, \cdot)$ \cite{bdfhp2009},
each $\mathcal{P}_{ab}$ can be expanded as an integral over an
element
$[z_{ab}]$ of $\CP^1$.
Here
$[z]$ denotes the equivalence class  in $\CP^1$ of a spinor $z$, that is, modulo rescaling by
$\C\setminus\{0\}$.
One obtains
\begin{align*}
A_v = c \int_{SL(2,\mathbb{C})^5} \delta(g_0) \left(\prod_a dg_a\right) \int_{(\mathbb{CP}^1)^{10}} \prod_{a<b} d{\mu}_{z_{ab}} e^\SEPRLorig
\end{align*}
where $c:=\frac{(1+\gamma^2)^5}{(\pi(1-i\gamma))^{10}}$ and
$d\mu_{z_{ab}} := \frac{-d_{j_{ab}}\Omega_{z_{ab}}}{\langle Z_{ab},Z_{ab} \rangle \langle Z_{ba},Z_{ba} \rangle}$,
with $\Omega_z := \frac{i}{2}(\epsilon_{AB}z^A dz^B)\wedge(\epsilon_{AB}\bar{z}^A d\bar{z}^B)$
and $\epsilon_{AB} =  \left(\begin{smallmatrix} 0 & 1 \\ -1 & 0 \end{smallmatrix}\right)$.
Here the action above is given by:
\begin{align*}
\SEPRLorig = \sum_{a<b} j_{ab}\left( \log\frac{\langle J\xi_{ab},Z_{ab}\rangle^2 \langle Z_{ba},\xi_{ba}\rangle^2}{\langle Z_{ab},Z_{ab} \rangle \langle Z_{ba},Z_{ba} \rangle} + i\gamma \log\frac{\langle Z_{ba},Z_{ba} \rangle}{\langle Z_{ab},Z_{ab} \rangle}\right)
\end{align*}	
with\footnote{Here we are using
the convention in \cite{bd2011} for defining $z_{ab}$. By a change of variables to $\tilde{z}_{ab}=Jz_{ab}$ we get the convention used in \cite{ez2015}.}
$Z_{ab} = g^{\dagger}_a z_{ab}$, $Z_{ba} = g^{\dagger}_b z_{ab}$.

\subsection{Proper vertex}
\label{subsect:proper}

As shown in \cite{bdfhp2009} the EPRL amplitude has two terms in the semi-classical limit. In
\cite{engle2011a, engle2012, ez2015, evz2015} the proper vertex amplitude was introduced and it was shown that its semi-classical limit contains only one term in which the Regge action appears with positive sign.
This amplitude is given by:

\begin{align}
\label{eqn:proporig}
A^{(+)}_v = \int_{SL(2,\mathbb{C})^5} \delta(g_0) \left(\prod_a dg_a\right) \prod_{a<b} \alpha(\mathcal{I}\psi_{ab}, g^{-1}_a g_b \mathcal{I} \Pi_{ba}(\{g_{ab}\}) \psi_{ba})
\end{align}
where the projector $\Pi_{ba}$ is
\begin{align*}
\Pi_{ba}[g_{ab}] = \Pi_{(0,\infty)}\left[\beta_{ba}[g]\text{tr}\left(\sigma_i g_{ba} g^{\dagger}_{ba}\right) L^i\right].
\end{align*}
Here $g_{ba} = g^{-1}_b g_a$, $L^i$ is the rotation generator in the spin $j_{ab}$ representation of $SU(2)$,
\begin{align*}
 \beta_{ba}[g] = \text{sgn} \left( \epsilon_{ijk} n_{bc}^i n_{bd}^j n_{be}^k \epsilon_{lmn} n_{ac}^l n_{ad}^m n_{ae}^n \right)
\end{align*}
where $\lbrace c,d,e \rbrace =\lbrace 0,\ldots,4 \rbrace \backslash \lbrace a,b \rbrace$, and
\begin{align*}
n_{ba}^i[g] = \frac{1}{2}\text{tr}(\sigma^i g_{ba} g^{\dagger}_{ba}).
\end{align*}
Again, the amplitude is independent of the choice of numbering of the tetrahedra,
so long as the numbering is consistent with the fixed orientation of the 4-simplex.
This follows from the \mbox{(skew-)symmetry} of $\alpha(\cdot, \cdot)$ together with the proof of theorem 6 in \cite{ez2015}.
Using theorem 6 in \cite{ez2015} and choosing the boundary states to be coherent states,
this amplitude becomes
\begin{align}
\label{propcoh}
A^{(+)}_v = \int_{SL(2,\mathbb{C})^5} \delta(g_0) \left(\prod_a dg_a\right) \prod_{a<b}  \alpha(\mathcal{I}\Pi_{ab}(\{g_{ab}\})C^{j_{ab}}_{\xi_{ab}}, g^{-1}_a g_b \mathcal{I} C^{j_{ab}}_{\xi_{ba}}).
\end{align}
In order to separate the projectors from the rest of the integrand,
a resolution of the identity  introducing spinor variables $\eta_{ab}$
\begin{equation}
\label{resolution}
\Pi_{ab}C_{\xi_{ab}}=\int d\tilde{\mu}_{\eta_{ab}}C_{\hat{\eta}_{ab}}( C_{\hat{\eta}_{ab}},\Pi_{ab}C_{\xi_{ab}})
\end{equation}
is inserted for each factor in (\ref{propcoh}),
where $(\cdot,\cdot)$ denotes the hermitian inner product on the relevant irrep of $\SU(2)$
to which its arguments belong and $\etan = \eta / \|\eta\| := \eta / \sqrt{\langle \eta, \eta \rangle}$.
This yields
\begin{equation*}
A^{(+)}_v = \int_{SL(2,\mathbb{C})^5} \delta(g_0) \left(\prod_a dg_a\right) \prod_{a<b} d\tilde{\mu}_{\eta_{ab}}\alpha(\mathcal{I} C^{j_{ab}}_{\hat{\eta}_{ab}}, g^{-1}_a g_b \mathcal{I} C^{j_{ab}}_{\xi_{ba}})(C_{\hat{\eta}_{ab}}^{j_{ab}}, \Pi_{ab}(\{g\})C_{\xi_{ab}}^{ab})
\end{equation*}
with $d\tilde{\mu}_{\eta_{ab}} =\frac{d_{j_{ab}}}{\pi}\Omega_{\etan_{ab}}$. Note $\Omega_{\etan} = \Omega_{\eta} / \|\eta\|^4$. Rewriting each inner product in terms of an integral over a spinor $z_{ab}$ as before,  we obtain the integral representation
\begin{align}
\label{eqn:propamp}
A^{(+)}_v = \int  \delta(g_0) \left(\prod_a dg_a\right) \int \prod_{a<b} d\tilde{\mu}_{\eta_{ab}} d{\mu}_{z_{ab}} e^{S_{\text{prop}}}
\end{align}
with action given by
\begin{align*}
 S_{\text{prop}} &=  S_{\text{EPRL}} + S_{\Pi} \\
S_{\text{EPRL}} &= \sum_{a<b} S^{ab}_{EPRL} := \sum_{a<b} \left(j_{ab} \log\frac{\langle J\hat{\eta}_{ab},Z_{ab}\rangle^2 \langle Z_{ba},\xi_{ba}\rangle^2}{\langle Z_{ab},Z_{ab} \rangle \langle Z_{ba},Z_{ba} \rangle} + i\gamma j_{ab}\log\frac{\langle Z_{ba},Z_{ba} \rangle}{\langle Z_{ab},Z_{ab} \rangle}\right) \\
S_{\Pi} &= \sum_{a<b} S^{ab}_\Pi := \sum_{a<b} \log(C_{\hat{\eta}_{ab}}, \Pi_{ab}(\{g_{ab}\})C_{\xi_{ab}})_{j_{ab}}
\end{align*}

\subsection{Graviton propagator}

The `graviton propagator' predicted by a given model of quantum gravity is the connected two-point correlation function of the
`metric operator' $q^{ab}(x)$. The (densitized) metric operator is defined as $q^{ab}(x):= \delta^{ij} (E_n^a)_i (E_n^b)_j$,
where point $x$ is identified with a node $n$ of the boundary spin network or equivalently with a tetrahedron $n$
of the triangulation,
 and $(E_n^a)_i$ is the flux operator through a surface dual to the triangle between the tetrahedra $a$ and $n$,
in the frame of tetrahedron $n$.\footnote{
At each $n$, $E_n^a \cdot E_n^b$ has 6 independent components since $a\neq b$ and
$a,b=\{0,...4\}\setminus \{n\}$.
Therefore, $E_n^a\cdot E_n^b$ can completely determine the 3-metric $h(n)_{\mu\nu}$ which is a $3 \times 3$
symmetric matrix that also has 6 independent components.
Explicitly, if $V_{na}^\mu$ denotes the tangent to the link $na$ at the node $n$, then $\{V_{na}^\mu\}$
spans the tangent space at $n$, and for $a\neq b$,
$h(n)_{\mu\nu} V_{na}^\mu V_{nb}^\nu = E_n^a \cdot E_n^b$
whereas for the diagonal components
\begin{align*}
h(n)_{\mu\nu}V_{na}^\mu V_{na}^\nu = h(n)_{\mu\nu} V_{na}^\mu \left(\sum_{c\neq a,n} \lambda_c V_{nc}^\nu\right)
= \sum_{c\neq a,n} \lambda_c h(n)_{\mu\nu} V_{na}^\mu V_{nc}^\nu
= \sum_{c\neq a,n} \lambda_c E^a_n \cdot E^b_n
\end{align*}
}
In \cite{bd2011} the graviton propagator for the Lorentzian EPRL spinfoam model with a single spinfoam vertex is calculated and its asymptotics is studied. We will calculate the spinfoam propagator in section \ref{sect:propagator} using the same framework as \cite{bd2011}
but with the proper vertex amplitude defined in section \ref{subsect:proper}.
Here we briefly recall the framework of \cite{bd2011}, generalized to an arbitrary spin-foam model with boundary states
matching loop quantum gravity, as well as recalling the results of \cite{bd2011}.

The spinfoam propagator is defined by the expression
\begin{equation}
\label{propagator}
G^{abcd}(x,y)=\frac{\left\langle W| q^{ab}(x)q^{cd}(y)|\Psi_o \right\rangle}{\langle W|\Psi_o \rangle} -\frac{\left\langle W|q^{ab}(x)|\Psi_o\right\rangle}{\left\langle W|\Psi_o\right\rangle}\frac{\left\langle W|q^{cd}(y)|\Psi_o\right\rangle}{\left\langle W|\Psi_o\right\rangle}
\end{equation}
where $\langle W|$ is the amplitude map of the spinfoam model in question
and $|\Psi_o\rangle$ is a Lorentzian semi-classical boundary state peaked both on intrinsic and extrinsic geometry and is constructed as follows.
The Lorentzian coherent spin network states $|j_{ab},\Upsilon_a(\{\vec{n}_{ab}\})\rangle$,
peaked on a choice of \textit{intrinsic} boundary geometry, are
 labeled by a set of  spins $j_{ab}$ and
Lorentzian coherent intertwiners $\Upsilon_a \in {\rm Inv} \otimes_{b:b\neq a}\Hil_{j_{ab}}$.
Each Lorentzian coherent intertwiner $\Upsilon_a(\{\vec{n}_{ab}\})$ is determined by a set of four
unit 3-vectors $\{\vec{n}_{ab}\}_{b:b\neq a}$ via
\begin{equation}
\label{lcs}
\Upsilon_{a}(\{\vec{n}_{ab}\})=\exp\left(-i\sum\limits_{a<b}\Theta_{ab}j_{ab}\right)
\int_{\SU(2)} dh \bigotimes_{b:b\neq a} h | j_{ab}, \vec{n}_{ab}\rangle
\end{equation}
where $|j_{ab}, \vec{n}_{ab} \rangle$ denote the coherent states $C_{\xi_ab}^j$ introduced in section \ref{sect:EPRL},
with the spinors $\xi_{ab}$ chosen such that (1.) they are unit, (2.) $n_{\xi_{ab}} = n_{ab}$, and
(3.) they satisfy the Regge state phase condition \cite{bdfhp2009}.
$e^{-i\sum_{a<b}\Theta_{ab}j_{ab}}$ is called Lorentzian-geometry phase with $\Theta_{ab}=\pi$ for thin wedges and  $\Theta_{ab}=0$ for thick wedges\footnote{All tetrahedra are assumed to be space-like so their normals are time-like. A wedge composed by two tetrahedra $a$ and $b$ that share the triangle $(ab)$ is called thick wedge if their outward timelike normals are both future-pointing or both past-pointing otherwise called thin wedge.}.
The Lorentzian semi-classical state $|\Psi_o\rangle$ is then defined by a superposition of Lorentzian coherent spin network states:
\begin{equation}
\label{eqn:psizero}
|\Psi_o\rangle=\sum\limits_{j_{ab}}\psi_{j_o,\phi_o}(j)|j_{ab},\Upsilon_a(\vec{n})\rangle
\end{equation}
with coefficients given by a Gaussian times a phase,
\begin{equation}
\label{eqn:psicoeff}
\psi_{j_o,\phi_o}(j)=\exp \left(-\sum\limits_{ab,cd}\gamma\alpha^{(ab)(cd)}\frac{j_{ab} - (j_{0})_{ab}}{\sqrt {(j_{0})_{ab}}}\frac{j_{cd} -( j_{0})_{cd}}{\sqrt {(j_{0})_{cd}}}-i\sum\limits_{ab}\gamma\phi_{0}^{ab}(j_{ab}-(j_{0})_{ab})\right)
\end{equation}
where $\phi_o^{ab}$ is the dihedral angle between tetrahedron $a$ and $b$,
representing the simplicial extrinsic curvature, $(j_o)_{ab}$ are the background spins on which the Gaussian is peaked,
and $\alpha^{(ab)(cd)}$ is a $10 \times 10$ matrix assumed to be complex with positive definite real part.

In work \cite{bd2011} the above framework was applied to the case of the EPRL model.
There it was shown that, in the limit $j_o \rightarrow \infty$ and $\gamma \rightarrow 0$
\cite{mp2011},
%
%
%
the EPRL propagator matches the result from linearized gravity.

\section{Graviton propagator of the proper vertex}
\label{sect:propagator}

\subsection{Exact expression}

Following a procedure similar to that in \cite{bd2011} we want to calculate the propagator for the proper vertex reviewed in
section \ref{subsect:proper}. In this case the propagator (\ref{propagator}) becomes
\begin{align*}
{G_{nm}^{abcd}}^+=\frac{\sum_{j}\psi(j)\langle W_+| E_n^a.E_n^bE_m^c.E_m^d|j,\Upsilon(\vec{n})\rangle}{\sum_{j}\psi(j)\langle W_+|j,\Upsilon(\vec{n})\rangle}-\frac{\sum_{j}\psi(j)\langle W_+| E_n^a.E_n^b|j,\Upsilon(\vec{n})\rangle}{\sum_{j}\psi(j)\langle W_+|j,\Upsilon(\vec{n})\rangle}\frac{\sum_{j}\psi(j)\langle W_+| E_m^c.E_m^d|j,\Upsilon(\vec{n})\rangle}{\sum_{j}\psi(j)\langle W_+|j,\Upsilon(\vec{n})\rangle}
%
%
\end{align*}
where (\ref{eqn:psizero}) are used as boundary states
and $\langle W_+ | : |\Psi \rangle \mapsto \langle W_+|\Psi\rangle$ is the proper vertex amplitude defined by (\ref{eqn:proporig}). 
The indices are restricted to $a,b,c,d\neq n,m$ and $n\neq m$.
As mentioned in the preliminaries,
the numbering of the tetrahedra is arbitrary up to the orientation it defines on the 4-simplex,
and the vertex amplitude is independent of the choice of such a numbering.
Using this arbitrariness, we can, without loss of generality, assume $a,b < n$ and $c,d < m$.
%
%
 $(E_n^a)^i$ is in the frame of tetrahedron $n$ and for $a<n$ acts on the right hand-side of the corresponding factor in
the expression (\ref{propcoh}) for the proper vertex amplitude.
Using this expression and the formulas (\ref{lcs}-\ref{eqn:psicoeff}) for the boundary state,
one obtains
\begin{align}\label{propprop1}
& \langle W_+ \mid  E_n^a.E_n^b\mid j,\Upsilon (\xi )  \rangle = \int_{SL(2,\mathbb{C})^5}e^{-i\sum_{ab}\Theta _{ab}j_{ab}}\left(\prod_{a=0}^{4}dg_{a}\right)\delta({g_0})\nonumber\\\nonumber
& \hspace{2cm} \left(\prod_{p< q,(pq)\neq (na),(pq)\neq (nb)}
\alpha (\mathcal{I}\Pi _{pq}(\{g\}) C_{\xi _{pq}}^{j_{cd}},g_{p}^{-1}g_q\mathcal{I}C_{\xi _{qp}}^{j_{pq}})\right)
\alpha (\mathcal{I}\Pi_{an}(\{g\}) C_{\xi _{an}}^{j_{an}},g_{a}^{-1}g_n\mathcal{I}{(E_{n}^a)}^iC_{\xi _{na}}^{j_{na}})\\
& \hspace{4.5cm} \alpha( \mathcal{I}\Pi_{bn}(\{g\}) C_{\xi _{bn}}^{j_{bn}},g_{b}^{-1}g_n\mathcal{I}{(E_{n}^b)}^iC_{\xi _{nb}}^{j_{nb}}).
\end{align}
In order to separate the projector in the formulas we insert the resolution of the identity (\ref{resolution})
in terms of new spinor variables $\eta_{cd}$:
\begin{align*}
\alpha (\mathcal{I}\Pi _{pq}(\{g\})C_{\xi _{pq}}^{j_{pq}},g_{p}^{-1}g_q\mathcal{I}C_{\xi _{qp}}^{j_{pq}})
=\int d{\tilde{\mu}}_{\eta_{pq}}\alpha(\mathcal{I}C_{\hat{\eta}_{pq}}^{j_{pq}},g_{p}^{-1}g_q\mathcal{I}C_{\xi _{qp}}^{j_{pq}})(C_{\hat{\eta}_{pq}}^{j_{pq}},\Pi_{pq}(\{g\})C_{\xi_{pq}}^{j_{pq}}) .
\end{align*}
and
\begin{align*}\alpha (\mathcal{I}\Pi_{an}(\{g\})C_{\xi _{an}}^{j_{an}},g_{a}^{-1}g_n{(E_{n}^a)}^i\mathcal{I}C_{\xi _{na}}^{j_{na}})=\int d{\tilde{\mu}_{\eta_{an}}}\alpha (\mathcal{I}C_{\hat{\eta}_{an}}^{j_{an}},g_{a}^{-1}g_n{(E_{n}^a)}^i\mathcal{I}C_{\xi _{na}}^{j_{na}})(C_{\hat{\eta}_{an}}^{j_{an}},\Pi_{an}(\{g\})C_{\xi _{an}}^{j_{an}})
\end{align*}
Therefore, (\ref{propprop1}) becomes
\begin{align*}
& \langle W_+ \mid E_n^a.E_n^b\mid j,\Upsilon (\xi )  \rangle=
\int_{SL(2,\mathbb{C})^5} e^{-i\sum_{ab}\Theta _{ab}j_{ab}}\left(\prod_{a=0}^{4}dg_{a}\right)\delta({g_0})\delta_{ij}\cdot\\
&\hspace{3cm}\cdot\prod_{p< q,(pq)\neq (na),(pq)\neq (nb)}\int d{\tilde{\mu}_{\eta_{pq}}}
\alpha(\mathcal{I}C_{\hat{\eta}_{pq}}^{j_{pq}},g_{p}^{-1}g_q\mathcal{I}C_{\xi _{qp}}^{j_{pq}})(C_{\hat{\eta} _{pq}}^{j_{pq}},\Pi_{pq}(\{g\})C_{\xi_{pq}}^{j_{pq}})\cdot\\
&\hspace{5.2cm}\cdot \int d{\tilde{\mu}_{\eta_{an}}}\alpha (\mathcal{I}C_{\hat{\eta} _{an}}^{j_{an}},g_{a}^{-1}g_n{(E_{n}^a)}^i\mathcal{I}C_{\xi _{na}}^{j_{na}})(C_{\hat{\eta}_{an}}^{j_{an}},\Pi_{an}(\{g\})C_{\xi_{an}}^{j_{an}})\cdot\\
&\hspace{5.2cm}\cdot \int d{\tilde{\mu}_{\eta_{bn}}}\alpha (\mathcal{I}C_{\hat{\eta} _{bn}}^{j_{bn}},g_{b}^{-1}g_n{(E_{n}^b)}^j\mathcal{I}C_{\xi _{nb}}^{j_{nb}})(C_{\hat{\eta}_{bn}}^{j_{bn}},\Pi_{bn}(\{g\})C_{\xi_{bn}}^{j_{bn}}).
\end{align*}
The insertions $Q_{an}^i=\alpha (\mathcal{I}C_{\hat{\eta} _{an}}^{j_{an}},g_{a}^{-1}g_n{(E_{n}^a)}^i\mathcal{I}C_{\xi _{na}}^{j_{na}})$ are evaluated in \cite{bd2011} as\footnote{Actually, the insertions that are calculated in \cite{bd2011} are $Q_{an}^i=\alpha (\mathcal{I}C_{\xi _{an}}^{j_{an}},g_{a}^{-1}g_n{(E_{n}^a)}^i\mathcal{I}C_{\xi _{na}}^{j_{na}})$. By replacing $\xi_{an}$ with $\hat{\eta}_{an}$ on the left-hand side we get our desired result.}
\begin{equation*}
Q_{an}^i=\int\frac{1}{\pi}d\mu_{{z}_{an}}K_{an}(A_{an}^i)
\end{equation*}
where
\begin{equation*}
K_{an}=\left (\frac{\langle Z_{na},Z_{na} \rangle}{\langle Z_{an},Z_{an} \rangle}\right )^{i\gamma j_{an}}\left (\frac{\langle J\hat{\eta}_{an},Z_{an} \rangle^2\langle Z_{na},\xi_{na} \rangle^2}{\langle Z_{an},Z_{an} \rangle\langle Z_{na},Z_{na} \rangle}\right )^{j_{an}}
\end{equation*}
and
\begin{equation*}
A_{an}^i\equiv\gamma j_{an}\frac{\langle \sigma^iZ_{na},\xi_{na}\rangle}{\langle Z_{na},\xi_{na}\rangle}.
\end{equation*}
Using these results one gets
\begin{align*}
\langle W_+ \mid E_n^a.E_n^b\mid j,\Upsilon (\xi )  \rangle&=
c\int_{SL(2,\mathbb{C})^5} \left(\prod_{a=0}^{4}dg_{a}\right)\delta({g_0}) \cdot\\
&\hspace{4.3cm}\cdot \int \left (\prod_{a\rq{}<b\rq{}}\frac{d_{j_{a\rq{}b\rq{}}}}{\pi}d\mu_{{z}_{a\rq{}b\rq{}}}d{\tilde{\mu}_{\eta_{a\rq{}b\rq{}}}}\right )q_n^{ab}e^{S_{prop}\rq{}}
\end{align*}
and
\begin{align*}
\langle W_+ \mid E_n^a.E_n^bE_m^c.E_m^d\mid j,\Upsilon (\xi )  \rangle&=c\int_{SL(2,\mathbb{C})^5}\left(\prod_{a=0}^{4}dg_{a}\right)\delta({g_0})\cdot\\
&\hspace{4.2cm}\cdot \int \left (\prod_{a\rq{}<b\rq{}}\frac{d_{j_{a\rq{}b\rq{}}}}{\pi}d\mu_{{z}_{a\rq{}b\rq{}}}d{\tilde{\mu}_{\eta_{a\rq{}b\rq{}}}}\right )q_n^{ab}q_n^{cd}e^{S_{prop}\rq{}}
\end{align*}
where $S_{prop}\rq{}$ and the $q$\rq{}s are respectively given by
\begin{align}
\nonumber
S_{prop}\rq{}&:=S_{prop}-i\sum_{a<b}\Theta_{ab}j_{ab}=S_{EPRL}+S_{\Pi}-i\sum_{a<b}\Theta_{ab}j_{ab}
\\
\label{eqn:qA}
q_n^{ab}&:=A_n^a.A_n^b
\end{align}
The $q_n^{ab}$\rq{}s turn out to be the same as in the EPRL case \cite{bd2011},
but here the EPRL action is replaced by the proper action.
So, using (\ref{eqn:psizero}), the final expression for the propagator becomes
\begin{align}
\label{prop2}
G_{nm}^{abcd+}&=\frac{\sum_j\int d^4gd^{10}\mu_{{z}_{}}d^{10}\tilde{\mu}_{\eta_{}}q_n^{ab}q_m^{cd}e^{S_{prop}^{tot}}}{\sum_j\int d^4gd^{10}\mu_{{z}_{}}d^{10}\tilde{\mu}_{\eta_{}}e^{ S_{prop}^{tot}}}\\
\nonumber
&\hspace{1in}-\frac{\sum_j\int d^4gd^{10}\mu_{{z}_{}}d^{10}\tilde{\mu}_{\eta_{}}q_n^{ab}e^{ S_{prop}^{tot}}}{\sum_j\int d^4gd^{10}\mu_{{z}_{}}d^{10}\tilde{\mu}_{\eta_{}}e^{ S_{prop}^{tot}}}\frac{\sum_j\int d^4gd^{10}\mu_{{z}_{}}d^{10}\tilde{\mu}_{\eta_{}}q_m^{cd}e^{ S_{prop}^{tot}}}{\sum_j\int d^4gd^{10}\mu_{{z}_{}}d^{10}\tilde{\mu}_{\eta_{}}e^{ S_{prop}^{tot}}}
\end{align}
where
\begin{align}
\label{totalaction}
S_{prop}^{tot}(j,g,z,\eta)&=
-\frac{1}{2}\sum\limits_{ab,cd}\gamma\alpha^{(ab)(cd)}\frac{j_{ab} - (j_{0})_{ab}}{\sqrt {(j_{0})_{ab}}}\frac{j_{cd} -( j_{0})_{cd}}{\sqrt {(j_{0})_{cd}}}\\
\nonumber
&\hspace{1in}-i\sum\limits_{ab}\gamma\phi_{0}^{ab}(j_{ab}-(j_{0})_{ab}) -i\sum_{a<b}\Theta_{ab}j_{ab}+S_{EPRL}+S_{\Pi}  .
\end{align}
Notice that the coefficients $\psi(j)$ are absorbed in to the total action.

\subsection{Asymptotic limit}\label{asymlim}

As in \cite{bd2011}, to pose the asymptotic limit, we scale all of the spins $(j_o)_{ab}$ uniformly, setting
$(j_o)_{ab} =: \lambda (\fixedj_o)_{ab}$, with $(\fixedj_o)_{ab}$ fixed, and consider the limit $\lambda \rightarrow \infty$.
In addition, in each sum in (\ref{prop2})
we perform a change of variables from  $j_{ab}$ to $\fixedj_{ab}:= j_{ab}/\lambda$.
$S^{tot}_{prop}$ then becomes
\begin{align*}
S^{tot}_{prop} = \lambda \left[\tilde{S}_\psi + \tilde{S}_{EPRL}\right]
+ \sum_{a<b} S^\Pi_{ab}
\end{align*}
where
\begin{align*}
 \tilde{S}_\psi &:= -\frac{1}{2}\sum\limits_{ab,cd}\gamma\alpha^{(ab)(cd)}\frac{\fixedj_{ab}
- (\fixedj_{0})_{ab}}{\sqrt {(\fixedj_{0})_{ab}}}\frac{\fixedj_{cd} -( \fixedj_{0})_{cd}}{\sqrt {(\fixedj_{0})_{cd}}}
-i\gamma \sum_{a<b} \phi^{ab}_o(\fixedj_{ab} - (\fixedj_o)_{ab}) - i \sum_{a<b} \Theta_{ab} \fixedj_{ab}
\end{align*}
and $\tilde{S}_{EPRL}:= \sum_{a<b} \tilde{S}_{EPRL}^{ab}$ is the same as $S_{EPRL}$ but with
$j_{ab}$ replaced by $\fixedj_{ab}$.
The projector part of the total action (\ref{totalaction}),
$\sum_{a<b}S_\Pi^{ab}$, is not linear in $\lambda$ but it \textit{is asymptotically linear}.
Specifically, let $\nu_{ab}$ be any unit spinor such that
\begin{align*}
\mathbf{n}_{\nu_{ab}} = \langle \nu_{ab}, \mathbf{\sigma} \nu_{ab} \rangle
= \beta_{ab}[g] \frac{\tr(g_{ab}g_{ab}^\dagger \mathbf{\sigma})}{|\tr(g_{ab}g_{ab}^\dagger \mathbf{\sigma})|},
\end{align*}
and let
\begin{align*}
x_{ab} := \langle \eta_{ab}, \nu_{ab} \rangle \langle \nu_{ab}, \xi_{ab}\rangle
\qquad y_{ab} := \langle \eta_{ab}, J \nu_{ab} \rangle \langle J \nu_{ab}, \xi_{ab}\rangle .
\end{align*}
%
%
%
%
Then we have \cite{evz2015}
\begin{align*}
\exp (S_\Pi^{ab}) \sim \tilde{f}_{ab}\exp(\lambda \tilde{S}_\Pi^{ab})
\end{align*}
where
\begin{align*}
\tilde{S}^{ab}_\Pi &:= \left\{ \begin{array}{ll}
2\fixedj \log (x_{ab}+y_{ab}) & \text{ if }|x_{ab}|>|y_{ab}|\text{ and }|x_{ab}+y_{ab}|^2\ge|4x_{ab}y_{ab}| \\
\fixedj \log (4 x_{ab} y_{ab}) & \text{ if }|x_{ab}|<|y_{ab}|\text{ or }|x_{ab}+y_{ab}|^2<|4x_{ab}y_{ab}|
\end{array} \right.\\
\tilde{f}_{ab}&:= \left\{ \begin{array}{ll} 1 & \text{ if }|x_{ab}|>|y_{ab}|\text{ and }|x_{ab}+y_{ab}|^2\ge|4x_{ab}y_{ab}| \\
\frac{1}{\sqrt{\pi \lambda \fixedj_{ab}}}
\frac{x_{ab}^{1-\alpha_{ab}} y_{ab}^{\alpha_{ab}}}{y_{ab}-x_{ab}} & \text{ if }|x_{ab}|<|y_{ab}|\text{ or }|x_{ab}+y_{ab}|^2<|4x_{ab}y_{ab}|
\end{array} \right.
\end{align*}
and
%
%
%
\begin{align*}
\alpha_{ab} = \left\{ \begin{array}{ll}
0 & \text{ if } j_{ab}=\lambda \fixedj_{ab} \text{ is integer} \\
1/2 & \text{ if } j_{ab}=\lambda \fixedj_{ab} \text{ is half integer.}
\end{array} \right.
\end{align*}
%
%
Lastly, let $\tilde{q}_n^{ab}$ be the same as $q_n^{ab}$, but with $j_{ab}$ replaced by $\fixedj_{ab}$,
so that $q_n^{ab} = \lambda^2 \tilde{q}_n^{ab}$.
With the above definitions, the asymptotic limit of (\ref{prop2}) is
\begin{align}
\label{proptilde}
\frac{1}{\lambda^4}G_{nm}^{abcd+}(\lambda)\sim&\frac{\sum_\fixedj\int d^4gd^{10}\mu_{{z}_{}}d^{10}\tilde{\mu}_{\eta}\tilde{f}
\tilde{q}_n^{ab}\tilde{q}_m^{cd}e^{\lambda \tilde{S}}}{\sum_\fixedj\int d^4gd^{10}\mu_{{z}_{}}d^{10}\tilde{\mu}_{\eta}\tilde{f}e^{ \lambda \tilde{S}}}\\
\nonumber
&\hspace{1in} -\frac{\sum_\fixedj\int d^4gd^{10}\mu_{{z}_{}}d^{10}\tilde{\mu}_{\eta}\tilde{f}
\tilde{q}_n^{ab}e^{ \lambda \tilde{S}}}{\sum_\fixedj\int d^4gd^{10}\mu_{{z}_{}}d^{10}\tilde{\mu}_{\eta}\tilde{f}e^{ \lambda \tilde{S}}}\frac{\sum_\fixedj\int d^4gd^{10}\mu_{{z}_{}}d^{10}\tilde{\mu}_{\eta}\tilde{f}
\tilde{q}_m^{cd}e^{ \lambda \tilde{S}}}{\sum_\fixedj \int d^4gd^{10}\mu_{{z}_{}}d^{10}\tilde{\mu}_{\eta}\tilde{f}e^{ \lambda \tilde{S}}}
\end{align}
where
\begin{align}
\label{eqn:Stilde}
\tilde{f}:= \prod_{a<b} \tilde{f}_{ab} \qquad \text{and} \qquad
\tilde{S}:= \tilde{S}_\psi +  \tilde{S}_{EPRL} + \tilde{S}_\Pi
\end{align}
with $\tilde{S}_\Pi = \sum_{a<b}\tilde{S}_\Pi^{ab}$.
Using the Euler-Maclaurin formula for each of the sums in (\ref{proptilde}), as well as the fact that each resulting integral is asymptotic to a non-zero polynomial in $\lambda$ (as will be seen below from the fact that each one has one critical point) one obtains, similar to \cite{bd2011},\footnote{In
the asymptotic limit, each sum over $j$'s in the expression (\ref{proptilde}) becomes
an integral with almost everywhere smooth integrand, so that stationary phase can later be applied.
Note, this is precisely where the phase convention for the coherent intertwiners $\Upsilon(\{\vec{n}_{ab}\})$
becomes important: Without this phase convention, the sum over $j$'s would not approach a continuous integral, so that
stationary phase would not be applicable \cite{bianchiprivate2015}.
}
\begin{align}
\label{propint}
\frac{1}{\lambda^4}G_{nm}^{abcd+}(\lambda)\sim&\frac{\int_j\int
d^4gd^{10}\mu_{{z}_{}}d^{10}\tilde{\mu}_{\eta}\tilde{f}\tilde{q}_n^{ab}\tilde{q}_m^{cd}e^{\lambda\tilde{S}}}{\int_j\int d^4gd^{10}\mu_{{z}_{}}d^{10}\tilde{\mu}_{\eta}\tilde{f}e^{ \lambda\tilde{S}}}\\
\nonumber
&\hspace{1in}-\frac{\int_j\int d^4gd^{10}\mu_{{z}_{}}d^{10}\tilde{\mu}_{\eta}\tilde{f}\tilde{q}_n^{ab}e^{ \lambda\tilde{S}}}{\int_j\int d^4 g d^{10}\mu_{{z}_{}}d^{10}\tilde{\mu}_{\eta}\tilde{f}e^{ \lambda\tilde{S}}}\frac{\int_j\int d^4 d^{10}\mu_{{z}_{}}d^{10}\tilde{\mu}_{\eta}\tilde{f}\tilde{q}_m^{cd}e^{ \lambda\tilde{S}}}{\int_j\int d^4 g d^{10}\mu_{{z}_{}}d^{10}\tilde{\mu}_{\eta}\tilde{f}e^{ \lambda\tilde{S}}} .
\end{align}
Critical point equations obtained from maximality, and by varying $z_{ab}$ and $\eta_{ab}$ are
\cite{evz2015}:
\begin{equation}\label{cp1}
\xi_{ba}=\frac{e^{i\phi_{ba}}}{\|Z_{ba}\|}g_{b}^{\dagger}z_{ab}=:\frac{e^{i\phi_{ba}}}{\|Z_{ba}\|}Z_{ba}
\end{equation}
\begin{equation}\label{cp2}
J\hat{\eta}_{ab}=\frac{e^{i\phi_{ab}}}{\|Z_{ab}\|}g_{a}^{\dagger}z_{ab}=:\frac{e^{i\phi_{ab}}}{\|Z_{ab}\|}Z_{ab}
\end{equation}
\begin{equation}\label{cp4}
\hat{\eta}_{ab}=e^{i\theta}\xi_{ab}
\end{equation}
\begin{equation}\label{cp5}
\beta_{ab}(\{g\}) n_{\xi_{ab}}^i \tr\left( g_{ab} g_{ab}^\dagger \sigma_i \right) > 0
\end{equation}
for all $a<b$,
for some set of phases $\phi_{ab}$, $\phi_{ba}$ and $\theta$. The variation of $\tilde{S}$
with respect to the group elements $g_a$ gives the closure condition on the boundary state, as in
\cite{bdfhp2009, evz2015}.

These equations determine $[z_{ab}], [\eta_{ab}]\in \CP^1$, and $g_a$ uniquely
\cite{evz2015}.\footnote{Note
that we have gauge-fixed the symmetry in the $g_a$'s with the factor
$\delta(g_0)$, as usual.
}
Furthermore, the final critical point equation, from varying $\fixedj_{ab}$, gives
\begin{align}
\label{jcrit}
\fixedj_{ab} = (\fixedj_o)_{ab}
\end{align}
so that \textit{there is exactly one critical point of this action}.
It follows that the asymptotic expression for each integral in
(\ref{propint}) will consist in only a single term.

The asymptotics can be evaluated using the extended stationary phase method.
As there is only one critical point, the asymptotic expression becomes \cite{bmp2009b,bd2011}
\begin{equation}
\label{gasym}
\frac{1}{\lambda^4}G^{abcd+}_{mn}(\lambda)= \frac{1}{\lambda}
(H^{-1}(x_o))^{ij}\frac{\partial \tilde{q}_n^{ab}(x_o)}{\partial x^i}\frac{\partial \tilde{q}_m^{cd}(x_o)}{\partial x^j}+\textrm{higher order terms}
\end{equation}
where $H$ is the Hessian of the asymptotic action $\tilde{S}$,
$x^i=\{\fixedj_{ab}, g_a, z_{ab}, \eta_{ab} \}$ and $x_o$ is the one critical point of $\tilde{S}$
(\ref{cp1}-\ref{jcrit}).

Note that if we define
\begin{align*}
\tilde{S}^o_\Pi:=
2\sum_{a<b}\fixedj_{ab} \log (x_{ab}+y_{ab}) = 2\sum_{a<b}\fixedj_{ab} \log \langle \hat{\eta}_{ab}, \xi_{ab} \rangle,
\end{align*}
then in a neighborhood of the critical point we have
\begin{align}
\label{Sfinal}
\tilde{S}= \Sfinal :=  \tilde{S}_\psi - i \sum_{a<b} \Theta_{ab} \fixedj_{ab} + \tilde{S}_{EPRL}
+ 2 \sum_{a<b} \fixedj_{ab} \log \langle \hat{\eta}_{ab}, \xi_{ab} \rangle
\end{align}
giving us an expression for the action that can be used to calculate the Hessian in the asymptotic formula.

%
%

\section{Hessian of the proper vertex: Calculation}
\label{sect:hess}

Calculating the Hessian of the asymptotic action is crucial in studying the asymptotics of the spinfoam propagator and also the asymptotics of the spinfoam vertex amplitude. In this section we evaluate the Hessian for the
proper action at the critical point satisfying (\ref{cp1}-\ref{jcrit}).

All second derivatives of the first three terms in (\ref{Sfinal}) with respect to $\fixedj_{ab}, g_a, z_{ab}$
were calculated in \cite{bd2011}.  To complete the calculation of the Hessian, it remains to calculate all second derivatives of
$\tilde{S}_{EPRL}$ involving $\eta_{ab}$, and all second derivatives of $\tilde{S}^o_\Pi$.
This is carried out in this section.

\subsection{Derivatives of $\tilde{S}_{EPRL}$ with respect to $\eta$ and one other variable}

To calculate the derivatives in the Hessian for use in the stationary phase theorem,
it is necessary to choose a section of $\CP^1$  (and coordinates thereon)  for each variable $\eta_{ab}$ and $z_{ab}$, as
well as a coordinate system on $SL(2,\C)$ for each variable $g_a$.
However, the specific form of the section for each $z_{ab}$ and the coordinate system for each $g_a$ turns out not to be needed for any of the results of this paper, so we leave them general.  Derivatives with respect to these general coordinates
we denote by $\delta_{z_{ab}}^i$, with $i=1,2$, and
$\delta_{g_a}^i$, with $i=1,\dots 6$ and the index $i$ sometimes suppressed.
%
%
For each variable $\eta_{ab}$, we introduce a section and complex coordinate $\etap_{ab}$,
with $\lvert\etap_{ab}\rvert \leq 1$,  defined by
\begin{align}
\label{zetadef}
\eta_{ab} = \sqrt{1-\lvert\etap_{ab}\rvert^2} \xi_{ab} + \etap_{ab} J\xi_{ab} .
\end{align}
The calculation of derivatives with respect to the $\eta$'s then reduces to taking holomorphic
and anti-holomorphic derivatives $\frac{\partial}{\partial\etap}$ and $\frac{\partial}{\partial\bar{\etap}}$.

We begin by noting
\begin{align*}
\frac{\partial}{\partial\etap_{ab}} \langle J\eta_{ab}, Z_{ab}\rangle &= - \frac{\bar{\etap}_{ab}}{2\sqrt{1-\etap_{ab}\bar{\etap}_{ab}}} \langle J\xi_{ab}, Z_{ab}\rangle
- \langle \xi_{ab}, Z_{ab}\rangle \\
\frac{\partial}{\partial\bar{\etap}_{ab}} \langle J\eta_{ab}, Z_{ab}\rangle &= - \frac{\etap_{ab}}{2\sqrt{1-\etap_{ab}\bar{\etap}_{ab}}} \langle J\xi_{ab}, Z_{ab}\rangle
\end{align*}
so that
\begin{align}
\label{eqn:zS}
\frac{\partial}{\partial\etap_{ab}} \tilde{S}_{EPRL}^{ab} &= - \fixedj_{ab} \frac{\bar{\etap}_{ab}}{\sqrt{1-\etap_{ab}\bar{\etap}_{ab}}} \frac{\langle J\xi_{ab}, Z_{ab}\rangle}{\langle J\eta_{ab}, Z_{ab}\rangle} - 2\fixedj_{ab} \frac{\langle \xi_{ab}, Z_{ab}\rangle}{\langle J\eta_{ab}, Z_{ab}\rangle} \\
\label{eqn:zbarS}
\frac{\partial}{\partial\overline{\etap}_{ab}} \tilde{S}_{EPRL}^{ab} &= - \fixedj_{ab} \frac{\etap_{ab}}{\sqrt{1-\etap_{ab}\bar{\etap}_{ab}}} \frac{\langle J\xi_{ab}, Z_{ab}\rangle}{\langle J\eta_{ab}, Z_{ab}\rangle}
\end{align}
Noting that, at the critical point $\langle \xi_{ab}, Z_{ab}\rangle = 0$ and $\etap_{ab} = 0$, we have
\begin{align*}
\left.\frac{\partial}{\partial\fixedj_{ab}} \frac{\partial}{\partial\etap_{ab}} \tilde{S}_{EPRL}^{ab}
\right|_\critpt
=
\left.\frac{\partial}{\partial\fixedj_{ab}} \frac{\partial}{\partial\bar{\etap}_{ab}} \tilde{S}_{EPRL}^{ab}
\right|_\critpt = 0
\end{align*}
Next, from (\ref{eqn:zS}-\ref{eqn:zbarS}), we calculate the mixed $\delta_g \delta_{\eta}$ derivatives:
\begin{align*}
\delta_{g_a} \frac{\partial}{\partial\etap_{ab}} \tilde{S}_{EPRL}^{ab}
&= -\fixedj_{ab} \frac{\overline{\zeta}_{ab}}{\sqrt{1-\zeta_{ab}\overline{\zeta}_{ab}}}
\delta_{g_a} \left( \frac{\langle J\xi_{ab}, Z_{ab}\rangle}{\langle J \eta_{ab}, Z_{ab}\rangle}\right) \\
& \hspace{1in} -2\fixedj_{ab} \left( \frac{\langle (\delta g_a) \xi_{ab}, z_{ab}\rangle}{\langle J\eta_{ab}, Z_{ab}\rangle}
- \frac{\langle \xi_{ab},Z_{ab}\rangle}{\langle J\eta_{ab}, Z_{ab}\rangle^2}
\langle (\delta g_a) J\eta_{ab}, Z_{ab}\rangle\right)
\\
\delta_{g_a} \frac{\partial}{\partial\bar{\etap}_{ab}} \tilde{S}_{EPRL}^{ab}
&= -\fixedj_{ab}\frac{\zeta_{ab}}{\sqrt{1-\zeta_{ab}\overline{\zeta}_{ab}}}
\delta_{g_a}\left( \frac{\langle J\xi_{ab}, Z_{ab}\rangle}{\langle J\eta_{ab}, Z_{ab}\rangle}\right)
\end{align*}
Evaluating at the critical point gives
\begin{align*}
\left.\frac{\partial}{\partial\etap_{ab}} \delta_{g_a} \tilde{S}_{\text{EPRL}}^{ab}\right|_\critpt
&= -2\fixedj_{ab} \langle (g_a^{-1} \delta g_a) \xi_{ab}, J \xi_{ab}\rangle \\
\left.\frac{\partial}{\partial\bar{\etap}_{ab}} \delta_{g_a} \tilde{S}_{\text{EPRL}}^{ab}\right|_\critpt&= 0
\end{align*}
The second derivative $\delta_{\eta_{ab}} \delta_g \tilde{S}^{ab}_{\text{EPRL}}$
for variations $\delta_g$ of any group elements other than $g_a$ is trivially zero.
The mixed derivatives $\delta_{\eta} \delta_{z}$ are
\begin{align*}
\delta_{z_{ab}} \frac{\partial}{\partial\etap_{ab}}  \tilde{S}^{ab}_{\text{EPRL}}
&= -\fixedj_{ab} \frac{\overline{\zeta}_{ab}}{\sqrt{1-\zeta_{ab}\overline{\zeta}_{ab}}}
\delta_{z_{ab}}\left(\frac{\langle J\xi_{ab}, Z_{ab}\rangle}{\langle J\eta_{ab}, Z_{ab}\rangle}\right) \\
&\hspace{1in} -2\fixedj_{ab}\left( \frac{\langle g_a \xi_{ab}, \delta z_{ab}\rangle}{\langle J\eta_{ab}, Z_{ab}\rangle}
- \frac{\langle \xi_{ab}, Z_{ab}\rangle}{\langle J\eta_{ab}, Z_{ab}\rangle^2}
\langle g_a J\eta_{ab}, \delta z_{ab}\rangle\right)
\\
 \delta_{z_{ab}} \frac{\partial}{\partial\bar{\etap}_{ab}} \tilde{S}^{ab}_{\text{EPRL}}
&= -\fixedj_{ab}\frac{\zeta_{ab}}{\sqrt{1-\zeta_{ab}\overline{\zeta}_{ab}}}
\delta_{z_{ab}}\left(\frac{\langle J\xi_{ab}, Z_{ab}\rangle}{\langle J\eta_{ab}, Z_{ab}\rangle} \right)
\end{align*}
Evaluating at the critical point yields
\begin{align*}
\left.\delta_{z_{ab}}\frac{\partial}{\partial\etap_{ab}}  \tilde{S}^{ab}_{\text{EPRL}}\right|_\critpt
&= -2(\fixedj_o)_{ab} \frac{\langle g_a \xi_{ab}, \delta z_{ab}\rangle}{\langle J\xi_{ab}, Z_{ab}\rangle} \\
\left.\delta_{z_{ab}}\frac{\partial}{\partial\bar{\etap}_{ab}}  \tilde{S}^{ab}_{\text{EPRL}}\right|_\critpt
&= 0
\end{align*}
The last to evaluate are the derivatives $\delta_{\eta} \delta_{\eta}$:
\begin{align*}
\frac{\partial}{\partial \zeta_{ab}} \frac{\partial}{\partial \zeta_{ab}} \tilde{S}_{EPRL}^{ab}
&= - \fixedj_{ab} \frac{\langle J \xi_{ab}, Z_{ab} \rangle}{\langle J\eta_{ab}, Z_{ab} \rangle}
\frac{\overline{\zeta}_{ab}^2}{2(1-\zeta_{ab}\overline{\zeta}_{ab})^{3/2}}\\
\frac{\partial}{\partial \zeta_{ab}} \frac{\partial}{\partial \overline{\zeta}_{ab}} \tilde{S}_{EPRL}^{ab}
&= - \fixedj_{ab} \frac{\langle J\xi_{ab}, Z_{ab} \rangle}{\langle J\eta_{ab}, Z_{ab} \rangle}
\left(\frac{1}{(1-\zeta_{ab} \overline{\zeta}_{ab})^{1/2}}
+ \frac{\zeta_{ab} \overline{\zeta}_{ab}}{2(1-\zeta_{ab} \overline{\zeta}_{ab})^{3/2}}\right)\\
\frac{\partial}{\partial \overline{\zeta}_{ab}}  \frac{\partial}{\partial \overline{\zeta}_{ab}}
\tilde{S}_{EPRL}^{ab} &= - \fixedj_{ab} \frac{\langle J\xi_{ab}, Z_{ab}\rangle}{\langle J\eta_{ab}, Z_{ab}\rangle}
\frac{\zeta_{ab}^2}{2(1-\zeta_{ab}\overline{\zeta}_{ab})^{3/2}}
\end{align*}
At the critical point $\etap_{ab} = 0$,
$\langle J \eta_{ab}, Z_{ab}\rangle = \langle J \xi_{ab}, Z_{ab}\rangle = 1$
and $\fixedj_{ab} = (\fixedj_o)_{ab}$,
so the only non-zero term is:
\begin{align*}
\left.\frac{\partial}{\partial\bar{\etap}_{ab}} \frac{\partial}{\partial\etap_{ab}}
\tilde{S}_{EPRL}^{ab}\right|_\critpt = -(\fixedj_o)_{ab}
\end{align*}

\subsection{Derivatives of $\tilde{S}^o_\Pi$}

We begin by writing $\tilde{S}^o_\Pi$ in terms of the coordinates $\zeta_{ab}$:
\begin{align*}
\tilde{S}^o_\Pi = 2 \sum_{a<b} \fixedj_{ab} \log \langle \hat{\eta}_{ab}, \xi_{ab} \rangle
= \sum_{a<b}  \fixedj_{ab} \log (1-\zeta_{ab} \overline{\zeta_{ab}})
\end{align*}
One then readily calculates
\begin{align*}
\frac{\partial}{\partial \zeta_{ab}} \frac{\partial}{\partial \fixedj_{ab}} \tilde{S}^o_\Pi
&= \frac{-\overline{\zeta}_{ab}}{1-\zeta_{ab}\overline{\zeta}_{ab}} \\
\frac{\partial}{\partial \overline{\zeta}_{ab}} \frac{\partial}{\partial\fixedj_{ab}} \tilde{S}^o_\Pi
&= \frac{-\zeta_{ab}}{1-\zeta_{ab}\overline{\zeta}_{ab}} \\
\frac{\partial}{\partial\zeta_{ab}} \frac{\partial}{\partial\zeta_{ab}} \tilde{S}^o_\Pi
&= \frac{-\fixedj_{ab} \overline{\zeta}_{ab}^2}{(1-\zeta_{ab}\overline{\zeta}_{ab})^2} \\
\frac{\partial}{\partial\overline{\zeta}_{ab}} \frac{\partial}{\partial\zeta_{ab}} \tilde{S}^o_\Pi
&= \frac{-\fixedj_{ab}}{(1-\zeta_{ab}\overline{\zeta}_{ab})^2} \\
\frac{\partial}{\partial\overline{\zeta}_{ab}} \frac{\partial}{\partial \overline{\zeta}_{ab}} \tilde{S}^o_\Pi
&=  \frac{-\fixedj_{ab}\zeta_{ab}^2}{(1-\zeta_{ab}\overline{\zeta}_{ab})^2}
\end{align*}
for all $a,b$, with all other second derivatives of $\tilde{S}^o_\Pi$ zero.
Evaluating these using the critical point equations $\zeta_{ab}=0$ and $\fixedj_{ab} = (\fixedj_o)_{ab}$,
only one second derivative of $\tilde{S}^o_\Pi$ is non-zero:
\begin{align*}
\left.\frac{\partial}{\partial\overline{\zeta}_{ab}} \frac{\partial}{\partial\zeta_{ab}}
\tilde{S}^o_\Pi\right|_\critpt
= \left.\frac{\partial}{\partial\zeta_{ab}} \frac{\partial}{\partial\overline{\zeta}_{ab}}
\tilde{S}^o_\Pi\right|_\critpt = -(\fixedj_o)_{ab} .
\end{align*}

\subsection{Hessian}
Bringing the above results together, we can write the Hessian as a
$(10+24+20+10+10)\times(10+24+20+10+10)$ matrix:
\begin{align*}
H(x_o) &:= S''(x_o) = \begin{pmatrix}
	Q_{\fixedj\fixedj}            & 0_{10\times24}        & 0_{10\times20}         & 0_{10\times10}  & 0_{10\times 10}\\
	0_{24\times10}    & H_{gg}                & H_{gz}                 & H_{g\etap} & 0_{24\times 10} \\
	0_{20\times10}    & H_{zg}                & H_{zz}                 & H_{z\etap} &  0_{20\times 10}\\
	0_{10\times10}    & H_{\etap g}             & H_{\etap z}              & 0_{10\times 10} & H_{\etap\bar{\etap}} \\
0_{10\times10}    & 0_{10\times 24}         &  0_{10\times 20}            & H_{\bar{\etap}\etap} & 0_{10\times 10}
	 \end{pmatrix}
\end{align*}
where
\begin{align*}
Q_{\fixedj\fixedj} &= Q^{\text{BD}}_{\fixedj\fixedj} \\
H_{gg} &= H^{\text{BD}}_{gg} \\
H_{gz} &= H^{\text{BD}}_{gz} \\
H_{zz} &= H^{\text{BD}}_{zz}
\end{align*}
are the Hessian components calculated by Bianchi and Ding
\cite{bd2011}. $H_{g\etap}$, $H_{z\etap}$, and $H_{\etap\bar{\etap}}$ are respectively given by
\begin{align*}
H^{g\etap}_{(ai)(ab)} &= -2(\fixedj_o)_{ab} \langle (g_a^{-1} \delta^i g_a) \xi_{ab}, J\xi_{ab} \rangle \\
H^{z\etap}_{(ab)(cd)} &= -2(\fixedj_o)_{ab}
\frac{\langle g_a \xi_{ab}, \delta z_{ab} \rangle}{\langle J\xi_{ab}, Z_{ab} \rangle}
\delta_{ac}\delta_{bd}\\
H^{\etap\bar{\etap}}_{(ab)(cd)} &= -2(\fixedj_o)_{ab} \delta_{ac}\delta_{bd} .
\end{align*}

\section{Graviton propagator and its $\gamma \rightarrow 0$ limit.}
\label{sect:gammazero}

According to (\ref{gasym}),
calculation of the graviton propagator requires us to invert the Hessian matrix calculated in the last section.
Doing this exactly is a non-trivial task.
However, as was done in the work \cite{bd2011}, we consider the $\gamma \rightarrow 0$ limit
\cite{mp2011},
in which case the Hessian matrix simplifies greatly.
Such a limit can be motivated by considering the limit in which,
as $\lambda$ becomes large, the eigenvalues of areas,
which are proportional to $\gamma \lambda$, stay the same, so that $\gamma \rightarrow 0$
at the same time.
%
%
This is a limit often considered \cite{bd2011, bmp2009b, mp2011, mp2011a, bojowald2001}.

 According to the results in \cite{bd2011} and the results in the Hessian section \ref{sect:hess},
 all the nonvanishing elements of
 the Hessian at critical points are independent of $\gamma$ except $Q_{jj}^{BD}$,
which is linear in $\gamma$. If we write the Hessian in the form:
\begin{equation*}
H(x_o)=\left ( \begin{array}{cc}
Q_{\fixedj\fixedj}^{BD}&0_{10\times 64}\\
0_{64\times 10}& H^R
\end{array}\right )
\end{equation*}
then its inverse is
\begin{equation}
\label{invHes}
(H(x_o))^{-1}=\left ( \begin{array}{cc}
\left(Q_{\fixedj\fixedj}^{BD}\right)^{-1}&0_{10\times 64}\\
0_{64\times 10}& (H^R)^{-1}
\end{array}\right ).
\end{equation}
The $\tilde{q}_n^{ab}$ calculated using the proper action in this paper, equation (\ref{eqn:qA}), turn out to be the same as the
$q_n^{ab}$ obtained in \cite{bd2011}.
%
Consequently their derivatives
$\delta_{\bar{z}_{an}} \tilde{q}_n^{ab}|_\crit$,
$\delta_{g_a} \tilde{q}_n^{ab}|_\crit$,
$\delta_{\fixedj_{cd}} \tilde{q}_n^{ab}|_\crit$ evaluated at critical points (\ref{cp1}-\ref{jcrit}) 
are the same and they scale as $\gamma^2$ and $\delta_{z_{an}}\tilde{q}_n^{ab}=0$.
Here we have two more sets of variables $\etap$ and $\bar{\etap}$,
but $\tilde{q}_n^{ab}$ does not depend on these variables: 
$\delta_{\etap}\tilde{q}_n^{ab}=\delta_{\bar{\etap}}\tilde{q}_n^{ab}=0$.
Using the derivatives above and the matrix (\ref{invHes}) in the asymptotic formula (\ref{gasym}) yields the same results as \cite{bd2011}
\begin{align}
G_{nm}^{abcd}=\lambda^3
\sum\limits_{p,q,r,s} \left(Q_{(pq)(rs)}^{BD}\right)^{-1}\frac{\partial \tilde{q}_n^{ab}}{\partial \fixedj_{pq}}
\frac{\partial \tilde{q}_m^{cd}}{\partial \fixedj_{rs}} + o(\gamma^3) + o(\lambda^3)
\end{align}
where $f=o(g)$ if $f/g$ vanishes in the limit $\lambda \rightarrow \infty$ and $\gamma \rightarrow 0$.
%
%
In this limit the propagator is asymptotic to
\begin{align*}
G_{nm}^{abcd}\sim \lambda^3 \sum\limits_{p,q,r,s} \left(Q_{(pq)(rs)}^{BD}\right)^{-1}\frac{\partial \tilde{q}_n^{ab}}{\partial \fixedj_{pq}}
\frac{\partial \tilde{q}_m^{cd}}{\partial \fixedj_{rs}}.
\end{align*}

\section{Coefficient in the asymptotics of the proper vertex}
\label{sect:coeff}

Lastly, we use the calculation of the Hessian to find the large spin limit of the proper vertex amplitude,
including the overall coefficient.
From (\ref{eqn:propamp}) we have
\begin{equation*}
A_{\nu}^+= c \int_{SL(2,\mathbb{C})}\left(\prod_{a=0}^{4}dg_{a}\right)\delta({g_0})\int_{(\mathbb{C}\mathbb{P}^1)^{10}}
\prod_{a<b}d\mu_{z_{ab}} \prod_{a<b} d\tilde{\mu}_{\eta_{ab}}e^{S_{prop}}.
\end{equation*}
As before, we set $j_{ab} = \lambda \fixedj_{ab}$ and consider the $\lambda \rightarrow \infty$ limit.
According to the argument in \cite{evz2015} and in section \ref{asymlim},
 in this limit the integrand can be replaced by its asymptotic form:
\begin{equation*}
A_{\nu}^+ \sim c \int_{SL(2,\mathbb{C})}\left(\prod_{a=0}^{4}dg_{a}\right)\delta({g_0})\int_{(\mathbb{C}\mathbb{P}^1)^{10}}
\left(\prod_{a<b}d\mu_{z_{ab}} d\tilde{\mu}_{\eta_{ab}}\right)\tilde{f}e^{\lambda \tilde{S}_{prop}}
\end{equation*}
where $\tilde{S}_{prop} := \tilde{S}_{EPRL} + \tilde{S}_\Pi$.
Using the fact that $\tilde{f}$ is constant and equal to 1 in a neighborhood of all critical points,
the stationary phase method gives us
\begin{align}
\label{propasym}
A^+_v \sim \left( \frac{2\pi}{\lambda}\right)^{32} \left( \det (-H^R(x_o))\right)^{-1/2}
\left(
\prod_{a=1}^4 \left[d\mu_{g_a}^{\text{Haar}}\right]
 \prod_{a<b} \left[d\mu_{z_{ab}}\right]
 \left[d\tilde{\mu}_{\eta_{ab}}\right]
\right)_{x=x_o} \cdot e^{\lambda \tilde{S}_{prop}(x_o)}
\end{align}
where $[d\mu]$ denotes the representation of a given measure $d\mu$ in the chosen coordinate system $\{x_i\}$
on the integration domain, in the sense that $d\mu = [d\mu] \prod_i dx_i$,
and where $x_o$ is the critical point.
We need to calculate the determinant of the Hessian $H^{R}$ of $\tilde{S}_{prop}$
with respect to $g_a$, $z_{ab}$, and $\eta_{ab}$, at $x_o$.
Since $\tilde{S} - \tilde{S}_{prop} = \tilde{S}_\psi$ is independent of these variables,
the derivatives of $\tilde{S}$ and $\tilde{S}_{prop}$ with respect to them are the same, so that
\begin{equation}
\label{Hprop}
H^{R}(x_o) =\left ( \begin{array}{ccccc}
 H_{gg}^{BD}&H_{gz}^{BD}&H_{g\bar{z}}^{BD}&H_{g\etap}&0_{24\times 10}\\
H_{zg}^{BD}&H_{zz}^{BD}&H_{z\bar{z}}^{BD}&H_{z\etap}&0_{10\times 10}\\
H_{\bar{z}g}^{BD}&H_{\bar{z}z}^{BD}&H_{\bar{z}\bar{z}}^{BD}&0_{10\times 10}&0_{10\times 10}\\
H_{\etap g}&H_{\etap z}&0_{10\times 10}&0_{10\times 10}&H_{\etap \bar{\etap}}\\
0_{10\times 10}&0_{10\times 10}&0_{10\times 10}&H_{\bar{\etap} \etap}&0_{10\times 10}
\end{array}\right ) .
\end{equation}
The Hessian relevant for the asymptotics of the original EPRL vertex amplitude is
\begin{equation*}
H^{EPRL}
= \left(  \begin{array}{ccc}
 H_{gg}^{BD}&H_{gz}^{BD}&H_{g\bar{z}}^{BD} \\
H_{zg}^{BD}&H_{zz}^{BD}&H_{z\bar{z}}^{BD}\\
H_{\bar{z}g}^{BD}&H_{\bar{z}z}^{BD}&H_{\bar{z}\bar{z}}^{BD} .
\end{array} \right)
\end{equation*}
Applying to (\ref{Hprop}) the usual formula \cite{powell2011} for the determinant of block matrices, together with
the skew symmetry of the determinant under exchange of columns, yields
%
%
\begin{equation}
\label{HR}
\det (-H^{R}(x_o))= (\det (-H_{\etap\bar{\etap}}))^2 \det (-H^{EPRL}(x_o))
\end{equation}
with
\begin{equation*}
\det (-H_{\etap\bar{\etap}})=\prod_{a<b}2 (\fixedj_o)_{ab}=2^{10}\prod_{a<b}(\fixedj_o)_{ab} .
\end{equation*}
In the asymptotic expression (\ref{propasym}), the correct branch of the square root of $\det(-H^{R}(x_o))$
must be deduced from the prescription described in \cite{hormander1983}.
This prescription is given in terms of real coordinates on the integration manifold:
in this case it turns out to consist in taking the product of the square root of each eigenvalue of
minus the Hessian $H$,
with branch cut along the negative real axis. (None of the eigenvalues of $-H$ lie on the negative real axis because the real part of $-H$ is always positive definite when using real coordinates.)  In our case, some of the coordinates are complex, and we have
calculated $-H^{R}(x_o)$ using holomorphic and anti-holomorphic derivatives.
By  using the relation between holomorphic and antiholomorphic derivatives,
and derivatives with respect to real and imaginary parts, one finds the correct square root of (\ref{HR})
to be
\begin{align*}
\sqrt{\det (-H^{R}(x_o))}= -2^{10}\left(\prod_{a<b}\fixedj_{ab}\right) \sqrt{\det (-H^{EPRL}(x_o))} .
\end{align*}
%
%
%

The projector part of the action vanishes at critical points so  $\tilde{S}_{prop}(x_o)=\tilde{S}_{EPRL}(x_0)$.
Furthermore, in terms of the coordinates $\zeta_{ab}$, $\overline{\zeta}_{ab}$, $d\tilde{\mu}_{\eta_{ab}}$ is given by
\begin{align*}
d\tilde{\mu}_{\eta_{ab}}
&= \frac{d_{j_{ab}}}{\pi} \frac{i}{2} \left(\epsilon_{AB}\eta_{ab}^A d\eta_{ab}^B\right)
\wedge \left(\epsilon_{AB}\overline{\eta}_{ab}^A d\overline{\eta}_{ab}^B\right)\\
&\criteq \frac{d_{j_{ab}}}{\pi} \frac{i}{2} d\zeta_{ab} \wedge d\overline{\zeta}_{ab}.
\end{align*}
Hence
\begin{align*}
\prod_{a<b} \left[ d\tilde{\mu}_{\eta_{ab}}\right]
\criteq \prod_{a<b} \left( \frac{i(2j_{ab}+1)}{2\pi}\right)
%
%
\sim -\left(\frac{\lambda}{\pi}\right)\pi^{10} \prod_{a<b} \fixedj_{ab}
\; \criteq \;  -\left(\frac{\lambda}{\pi}\right)\pi^{10} \prod_{a<b} (\fixedj_o)_{ab} .
\end{align*}
Therefore (\ref{propasym}) becomes
\begin{align*}
A^+_v \sim \left( \frac{2\pi}{\lambda}\right)^{22} \left( \det (-H_{EPRL}(x_o))\right)^{-1/2}
\left(
\prod_{a=1}^4 \left[d\mu_{g_a}^{\text{Haar}}\right]
 \prod_{a<b} \left[d\mu_{z_{ab}}\right]
\right)_{x=x_o} \cdot e^{\lambda \tilde{S}_{EPRL}(x_o)}
\end{align*}
which is identical to the term in the asymptotics of the EPRL vertex amplitude corresponding to the
Einstein-Hilbert sector, including overall coefficient.

\textit{Remark:}
In the above argument we have used the asymptotic action in the stationary phase formula.
If one instead uses the exact action, the answer does not change, as both of these actions, as well as their Hessians,
are equal at the critical point, as is shown in appendix \ref{app:exact}.

\section{Discussion}
In this paper we have studied the connected two-point correlation function for metric operators ---
also called the graviton propagator --- in the recently introduced proper spin-foam model.
To perform the analysis, we defined the correlation function in the boundary amplitude formalism using
the same semi-classical boundary state as in \cite{bd2011}, and evaluating to lowest order in the vertex expansion.
We then considered the asymptotic limit of large spins $j\rightarrow\infty$ up to the leading order. This limit was investigated using extended stationary phase methods which necessitated calculation of the Hessian of the proper vertex. This calculation comprised the main task of this work. Then we examined the limit of $\gamma\rightarrow0$ and found that in this limit our result exactly matches the graviton propagator of the EPRL model calculated in \cite{bd2011}, and therefore also matches
the graviton propagator calculated using linearized quantum gravity.
This provides one check on the validity of the proper spin-foam model in the weak curvature regime.

Calculation of the Hessian also allowed us to complete the asymptotic analysis of the proper vertex \cite{evz2015}.
Specifically we were able to evaluate the overall coefficient arising in the semi-classical limit. This factor matches exactly the factor appearing in the semi-classical limit of the EPRL vertex \cite{bdfhp2009}
in front of the term corresponding to the Einstein-Hilbert sector.

In this work we estimated the two-point correlation function for a single spin-foam vertex. The obvious direction for future work is to extend the analysis using the proper vertex to the case of multiple vertices. Since the proper vertex amplitude has the correct semi-classical limit, we believe this calculation will again be consistent with linearized quantum gravity in the appropriate regime. 
Specifically, we expect the proper vertex to remove unphysical contributions to the semi-classical limit which would otherwise be
generated by the extra terms in the EPRL vertex.

\section*{Acknowledgements}

The authors thank Chris Beetle for helpful suggestions used in appendix \ref{app:exact}.
This work was supported in part by the National Science Foundation through grants  PHY-1205968
and PHY-1505490.

\appendix

\section{Equality of Hessians of exact and asymptotic actions of the proper vertex}
\label{app:exact}

In this appendix we show that the second derivatives of $S_\Pi$ with respect to $g_a$ and $\eta_{ab}$
are equal to those of $\lambda \tilde{S}_\Pi$ when evaluated at the critical point.
This will establish that the Hessian of the exact and asymptotic actions, $S_{\text{prop}}$
and $\lambda (\tilde{S}_{\text{EPRL}} + \tilde{S}_\Pi)$, with respect to $g_a$ and $\eta_{ab}$,
are equal at the critical point.

\subsection{Calculation of derivatives of the projector action with respect to $g$}
\label{subsect:g}

The projector action is given by
\begin{equation*}
S_\Pi [\{g_a, \eta_{ab}\}] = \sum_{a<b} \log
\langle j_{ab}, \eta_{ab}| \Pi_{ab}\left(\{g_a\}\right)|j_{ab}, \xi_{ab} \rangle
\end{equation*}
where we use the bra-ket notation $|j_{ab}, \xi_{ab} \rangle$ for the coherent state
$C^{j_{ab}}_{\xi_{ab}}$

The calculation of these derivatives requires varying the projector $\Pi_{ab}$ with respect to the group variables. 
To do that we write the spectral projector as:
\begin{align*}
\Pi_{ab} = \Pi_{(0,\infty)}\left[\beta_{ab}(g)n_{ab}(g) \cdot L\right]
\end{align*}
where the normal $n$ is defined as:
\begin{align*}
n^i_{ab}(g) = \frac{\text{tr}\left(\sigma^i g_{ab} g^{\dagger}_{ab}\right)}{\left\lvert \text{tr}\left(\sigma^i g_{ab} g^{\dagger}_{ab}\right) \right\rvert}
\end{align*}
with $g_{ab} = g^{-1}_a g_b$.

Since $\beta_{ab}(g)$ is a sign function, its variation at the critical point does not contribute to the variation of the projection operator.
Let $\delta$ be any variation with respect to the group elements and let $g(t)$ be a path in $SL(2,\C)^5$
such that $\delta  = \frac{dg}{dt}|_{t=0}$.
Note that
\begin{align*}
\Pi_{ab}(g(t)) = r_{ab}(t) \Pi_{ab}(g(0)) r^{-1}_{ab}(t)
\end{align*}
for some $r_{ab}(t) \in SU(2)$.
It follows that the variation of the projector is given by:
\begin{align}
\label{eqn:projgvar}
\delta \Pi_{ab}(g) = \frac{d}{dt} \Pi_{ab}(g(t))\Big|_{t=0} = [\delta r, \Pi_{ab}(g(0))] =: i [v_i L^i, \Pi_{ab}(g(0)) ]
\end{align}
for some $v_i$.  Let $\delta'$ be any other variation with respect to the group elements with
$\delta' \Pi_{ab} =: i[v_i' L^i, \Pi_{ab}]$.
Then the second variation of the projector action is:
\begin{align*}
\delta' \delta S_\Pi = - &\sum_{a<b} \frac{\left\langle j_{ab},\eta_{ab} \lvert [v_i L^i, [v'_j L^j, \Pi_{ab}(g(0))]] \rvert j_{ab}, \xi_{ab} \right\rangle}{\left\langle j_{ab},\eta_{ab} \lvert \Pi_{ab}(g) \rvert j_{ab}, \xi_{ab} \right\rangle} \\ + &\sum_{a<b} \frac{\left\langle j_{ab},\eta_{ab} \lvert [v_i L^i, \Pi_{ab}(g(0))] \rvert j_{ab}, \xi_{ab} \right\rangle \left\langle j_{ab},\eta_{ab} \lvert [v'_j L^j, \Pi_{ab}(g(0))] \rvert j_{ab}, \xi_{ab} \right\rangle}{\left\langle j_{ab},\eta_{ab} \lvert \Pi_{ab}(g) \rvert j_{ab}, \xi_{ab} \right\rangle^2}
\end{align*}
At the critical point, however, $[\eta_{ab}] = [\xi_{ab}]$ and $\Pi_{ab}(g)$ acts as the identity on the corresponding coherent
state, so
\begin{align}
\label{eqn:projfirst}
\left\langle j_{ab},\eta_{ab} \lvert [v_i L^i, \Pi_{ab}(g(0))] \rvert j_{ab}, \xi_{ab} \right\rangle\bigg|_\critpt = 0
\end{align}
Next, define for each spinor $\xi$ the $SU(2)$ element \cite{bdfhp2009}
\begin{align*}
g(\xi) := \left(\begin{array}{cc}
\hat{\xi}_0 & -\overline{\hat{\xi}_1}\\
\hat{\xi}_1 & \overline{\hat{\xi}_0}
\end{array}\right),
\end{align*}
where $\hat{\xi}:= \xi/||\xi||$, and define
\begin{align*}
|\xi; j, m\rangle := g(\xi) |j,m\rangle
\end{align*}
where $|j,m\rangle$, as usual, are the states satisfying $J_z |j,m\rangle = m |j,m\rangle$.
Note that
\begin{align*}
(v\cdot L)\left|j_{ab},\xi_{ab}\right\rangle
= (v \cdot L) |\xi_{ab}; j_{ab}, j_{ab}\rangle
= \alpha_v \left|\xi_{ab}; j_{ab}, j_{ab}\right\rangle + \beta_v \left|\xi_{ab}; j_{ab}, j_{ab}-1\right\rangle
\end{align*}
for some $\alpha_v$ and $\beta_v$ and likewise for $v' \cdot L$.
Furthermore,  at the critical point, for $j>1$ (which holds in the limit which interests us),
$\Pi_{ab}(g)$ acts as the identity on both  $\left|\xi_{ab}; j_{ab}, j_{ab}\right\rangle$
and $\left|\xi_{ab}; j_{ab}, j_{ab}-1\right\rangle$.  These facts imply
\begin{align*}
\langle j_{ab},\eta_{ab} \lvert [v_i L^i, [v'_j L^j, \Pi_{ab}(g(0))]] \rvert j_{ab}, \xi_{ab} \rangle\bigg|_\critpt = 0 .
\end{align*}
We therefore conclude that
\begin{align*}
\delta' \delta S_\Pi\bigg|_\critpt = 0
\end{align*}
for all variations $\delta, \delta'$ of the $g_a$'s.

\subsection{Calculation of derivatives of the projector action with respect to $\eta$}
\label{subsect:eta}

We begin with the mixed $\delta_g \delta_{\eta}$ derivatives. From (\ref{eqn:projgvar}),
\begin{align*}
\delta_g S_\Pi^{ab} = i \frac{\left\langle j_{ab}, \eta_{ab} \Big| \left[v_i L^i, \Pi_{ab}\right] \Big| j_{ab}, \xi_{ab}\right\rangle}{\left\langle j_{ab}, \eta_{ab} \lvert \Pi_{ab} \rvert j_{ab}, \xi_{ab}\right\rangle}
\end{align*}
for some $v^i \in \mathbb{R}^3$.  Because the coherent states $|j_{ab}, \eta_{ab}\rangle$, for any fixed $j_{ab}$,
are all related by the action of some $SU(2)$ element, it follows that, for any real variation $\delta_\eta$ of $\eta$,  there exist $(v')^i \in \mathbb{R}^3$ such that
%
%
\begin{align*}
\delta_\eta |j_{ab}, \eta_{ab}\rangle &= i (v' \cdot L) |j_{ab}, \eta_{ab}\rangle
\end{align*}
so that
\begin{align*}
\delta_\eta \delta_g S_\Pi^{ab}
=
\frac{\left\langle j_{ab}, \eta_{ab} \left| v'\cdot L
\left[v \cdot  L, \Pi_{ab}\right] \right| j_{ab}, \xi_{ab}\right\rangle}
{\left\langle j_{ab}, \eta_{ab} \lvert \Pi_{ab}
\rvert j_{ab}, \xi_{ab}\right\rangle}
-
\frac{\left\langle j_{ab}, \eta_{ab} \left|
\left[v \cdot  L, \Pi_{ab}\right] \right| j_{ab}, \xi_{ab}\right\rangle
\left\langle j_{ab}, \eta_{ab} \left| v'\cdot L \right| j_{ab}, \xi_{ab}\right\rangle
}{\left\langle j_{ab}, \eta_{ab} \lvert \Pi_{ab}
\rvert j_{ab}, \xi_{ab}\right\rangle^2}
.
\end{align*}
Using the same arguments as in \ref{subsect:g}, we conclude that, at the critical point,
\begin{align*}
\delta_\eta \delta_g S_\Pi\bigg|_\critpt = 0
\end{align*}
for all variations $\delta_\eta, \delta_g$ of the $\eta$'s and $g$'s.

Finally, to calculate $\delta_{\eta}' \delta_{\eta} S_\Pi$, we will need the explicit section of
$\CP^1$ and coordinate $\zeta_{ab}$ thereon
introduced in equation (\ref{zetadef}). First, recall that \cite{bdfgh2009}
\begin{align}
\label{tenscoh}
| j_{ab}, \eta_{ab} \rangle &= \otimes^{2j_{ab}} | \eta_{ab} \rangle .
\end{align}
 Upon taking derivatives of (\ref{zetadef}) and evaluating at the critical point
($\zeta_{ab} = 0$, $\eta_{ab} = \xi_{ab}$), we find that the first derivatives and only non-zero second derivatives are
\begin{align*}
\left.\frac{\partial}{\partial \zeta_{ab}} \eta_{ab}(\zeta)\right|_\critpt &= J \xi_{ab} \\
\left.\frac{\partial}{\partial \overline{\zeta}_{ab}} \eta_{ab}(\zeta)\right|_\critpt &= 0 \\
\left.\frac{\partial}{\partial \overline{\zeta}_{ab}} \frac{\partial}{\partial \zeta_{ab}} \eta_{ab}(\zeta)
\right|_\critpt
&= -\frac{1}{2} \xi_{ab}
\end{align*}
with adjoints
\begin{align*}
\left.\frac{\partial}{\partial \overline{\zeta}_{ab}} \langle \eta_{ab} | \right|_\critpt &= \langle J \xi_{ab}| \\
\left.\frac{\partial}{\partial \zeta_{ab}} \langle \eta_{ab}| \right|_\critpt &= 0 \\
\left.\frac{\partial}{\partial \overline{\zeta}_{ab}} \frac{\partial}{\partial \zeta_{ab}} \langle \eta_{ab}|
\right|_\critpt
&= -\frac{1}{2} \langle \xi_{ab} | .
\end{align*}
These relations may then be used to evaluate the second derivatives of
\begin{align*}
S^\Pi = \sum_{a<b} \log \langle  j_{ab}, \eta_{ab}  | \Pi_{ab}(g) | j_{ab}, \xi_{ab}\rangle
\end{align*}
at the critical point.
Additionally using the fact that $\Pi_{ab}(g)$ acts on $|j_{ab}, \xi \rangle$ as the identity at the critical point,
the form (\ref{tenscoh}) for coherent states,
as well as the orthogonality of $\xi_{ab}$ and $J \xi_{ab}$,
one finds the only non-zero second derivatives of $S^\Pi$ at the critical point to be
\begin{align*}
\frac{\partial}{\partial \overline{\zeta}_{ab}} \frac{\partial}{\partial \zeta_{ab}}
S^\Pi = - j_{ab} = - \lambda \fixedj_{ab}.
\end{align*}


%
%

\end{document}